\documentclass[12pt,a4paper,dvips,onecolumn]{article}
\usepackage{glas}
\usepackage{epsfig}
\usepackage{graphicx,graphics}
\DeclareGraphicsExtensions{.eps .ps}                            
\begin{document}
\newcounter{DSN}
\setcounter{DSN}{0}

\begin{titlepage}{GLAS-PPE/1999-22}{Dec 1999}
\title{\large A Prototype RICH Detector Using
 Multi-Anode Photo Multiplier Tubes and Hybrid Photo-Diodes}
%\author{\ Authors et.al}
%\begin{Authlist}
\author{
E.~Albrecht\Iref{a2},
G.~Barber\Iref{a4}, 
J.H.~Bibby\Iref{a5},
N.H.~Brook\Iref{a3},
G.~Doucas\Iref{a5}, \\
A.~Duane\Iref{a4},
S.~Easo\Iref{a3},
L.~Eklund\Iref{a2},
M.~French\Iref{a6},
V.~Gibson\Iref{a1}, \\
T.~Gys\Iref{a2},
A.W.~Halley\Iref{a2},\Iref{a3},
N.~Harnew\Iref{a5},
M.~John\Iref{a4},
D.~Piedigrossi\Iref{a2}, \\
J.~Rademacker\Iref{a5},
B.~Simmons\Iref{a4},
N.~Smale\Iref{a5},
P.~Teixeira-Dias\Iref{a3},
L.~Toudup\Iref{a4}, \\
D.~Websdale\Iref{a4},
G.~Wilkinson\Iref{a5},
S.A.~Wotton\Iref{a1}.
}
%\end{Authlist}
\Instfoot{a1}{\it 
University of Cambridge, Cavendish Laboratory, 
Madingley Road, Cambridge CB3 0HE, UK.}
\Instfoot{a2}{\it
CERN, EP Division, 1211 Geneva 23, Switzerland.}
\Instfoot{a3}{\it
University of Glasgow, Department of Physics, 
Glasgow G12 8QQ, UK. }
\Instfoot{a4}{\it
Imperial College of Science Technology \& Medicine,
Blackett Laboratory, Prince Consort Road, London SW7 2AZ, UK.}
\Instfoot{a5}{\it
University of Oxford, Department of Nuclear Physics, Keble 
Road, Oxford OX1~3RH, UK.} 
\Instfoot{a6}{\it
Rutherford Appleton Laboratory, Chilton, Didcot, Oxon OX11 0QX, UK.}
%\date{\small 17-November-1998}
\begin {abstract}
  The performance of a prototype Ring Imaging 
  Cherenkov Detector is studied using a charged particle beam. 
   The detector performance, using $CF_{4}$ 
    and air as radiators, is described. Cherenkov angle precision 
    and photoelectron yield using hybrid photo-diodes and 
    multi-anode PMTs agree with simulations and are assessed
    in terms of the requirements of the LHCb experiment.
\end {abstract}
\end{titlepage}
%\maketitle
%\twocolumn
\sloppy
\section{\large Introduction}
 This paper reports results from a prototype Ring Imaging Cherenkov
(RICH) counter and compares the performance of 
Multi- \mbox{Anode} Photomultiplier tubes (MAPMT)
and two types of Hybrid Photo-diode  Detectors (HPD) for detecting
the Cherenkov photons. The experimental arrangement represents 
a prototype of the downstream RICH detector
of the LHCb experiment \cite {lhcb_proposal} at CERN.

 The LHCb experiment will make precision measurements of CP
asymmetries in B decays. Particle identification by the RICH detectors
is an important tool and an essential component of LHCb. 
For example, separating pions and kaons using the RICH suppresses 
backgrounds coming from  
 $B_{d}^{0}\rightarrow K^{+}\pi^{-}$, $B_{s}^{0}\rightarrow K^{+}\pi^{-}$
and $B_{s}^{0}\rightarrow K^{+} K^{-}$ when selecting
 $B_{d}^{0}\rightarrow\pi^{+}\pi^{-}$ decays, and backgrounds coming from
\mbox {$B_{s}\rightarrow D_{s}^{\pm}\pi^{\mp}$} when selecting the 
$B_{s}\rightarrow D_{s}^{\pm}K^{\mp}$ decay mode.

LHCb has two RICH detectors. Together they cover
polar angles from 10 to 330 mrad. The upstream detector, RICH1,
uses aerogel and $C_{4}F_{10}$ radiators to identify  
particles with momenta from 1 to 65 GeV/{\it c}. 
The downstream detector, RICH2, has 180 cm of 
$CF_{4}$ radiator and identifies particles with 
momenta up to 150 GeV/{\it c}. It uses a spherical focusing mirror 
with a radius of curvature of 820 cm which is tilted by 370 mrad 
to bring the image out of the acceptance of the spectrometer. 
A flat mirror then reflects this image onto the photodetector plane. 
For tracks with $\beta \simeq 1$, RICH2 
is expected to detect about 30 photoelectrons \cite {lhcb_proposal}.
 
 The LHCb collaboration intends to use arrays of photodetectors
 with a sensitive granularity of $2.5 \rm{mm} \times 2.5 \rm{mm}$  
covering an area of $2.9 \rm {m^{2}}$ with a total of 340,000
channels, to detect the Cherenkov photons in both RICH detectors. 
These photodetectors are expected to cover an active area of at 
least 70\% of the detector plane. Current 
commercially available \mbox{devices}\footnote { Commercial 
HPD devices from Delft Electronische Producten (DEP), 
The Netherlands, Commercial
MAPMT devices from Hamamatsu Photonics, Japan. } have
\mbox{inadequate} coverage of the active area 
and their performance at LHC speeds remains to be proven. 
The beam tests described here used prototypes of 
three of the new photodetector designs that have been
proposed for LHCb.

The results from the LHCb  RICH1 prototype 
detector tests carried out during 1997 are reported in an accompanying 
publication \cite {t_paper2}.
The data used in this paper were collected during the summer 
and autumn of 1998 at the CERN SPS facility. 
The main goals of these RICH2 prototype studies are:
\begin{itemize}
 \item[$\bullet$]{ To test the performance of the $CF_{4}$ radiator, using
  the full-scale optical layout of RICH2,}
 \item[$\bullet$]{ To test the performance of the photodetectors 
     using the RICH2 geometry by measuring the Cherenkov angle 
     resolution and photoelectron yields.}
\end {itemize}
 
 Section 2 of this paper describes the main features of the 
test beam setup. \mbox{Section 3} describes the simulation of the 
experiment and is followed by a discussion of the
photoelectron yields and Cherenkov angle \mbox{resolution} measurements for 
each of the photodetectors. Finally a summary is given in \mbox{Section 6},
with plans for future work.

\section{\large Experimental Setup}
\begin{figure}
\begin{center}
\epsfig{file=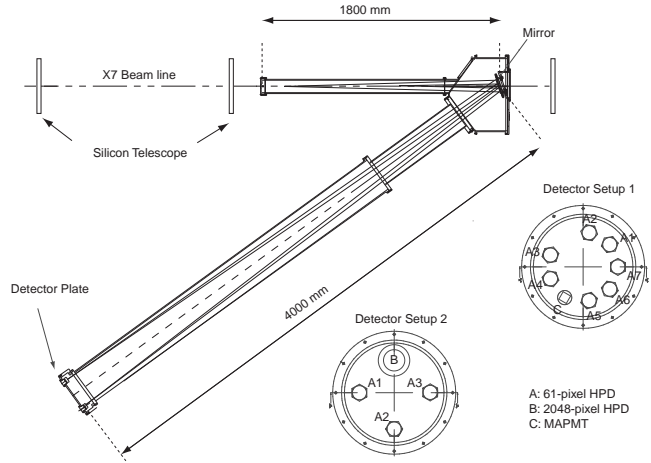,width=0.5\textwidth}
\end{center}
\caption{\label{schematic} Schematic diagram of the RICH2 test beam setup
  and two \mbox{photodetector} configurations.}
\end{figure}

   The setup included scintillators and 
a silicon telescope which defined and measured the direction of incident 
charged particles, a radiator for the production of Cherenkov photons,
a mirror for focusing these photons, photodetectors and 
the data acquisition system. A brief description of these 
components is given below, and a more complete 
\mbox{description} of the experimental setup can be 
found in \cite {t_paper1}. The photodetectors were mounted on a 
plate \mbox{customised} for particular detector configurations. 
A schematic diagram of the setup is shown in Figure \ref{schematic}.
\subsection{Beam line}
  The experimental setup was mounted in the CERN X7 beam line. 
 The beam was tuned to provide negative particles (mainly pions)
with momenta between \mbox{10 and 120 GeV/{\it c}}. The precision of the 
beam momentum for a given setting ($\delta$p/p) was better than 1\%. 
Readout of the detectors was triggered by the passage 
of beam particles which 
produced time-correlated signals from two pairs of scintillation counters 
placed  8 metres apart along the beam line. The beam size 
was $20\times 20 \rm {mm^{2}}$ as defined by the smaller of these counters.

\subsection{Beam Trajectory Measurement}

 The input beam direction and position were measured using a 
silicon telescope consisting of three planes of pixel detectors. 
Each of these planes has a $22\times 22$ array of silicon pixels 
with dimensions $1.3\rm {mm} \times 1.3 \rm {mm} $. Two of the planes were 
placed upstream of the radiator and the third one downstream of 
the mirror. The first and third planes were separated by 8 metres.

Using the silicon telescope, the beam divergence was measured 
to be typically 0.3 mrad and 0.1 mrad in the horizontal and 
vertical planes respectively. 
\subsection {The RICH Detector}
\begin{figure}
\begin{center}
\epsfig{file=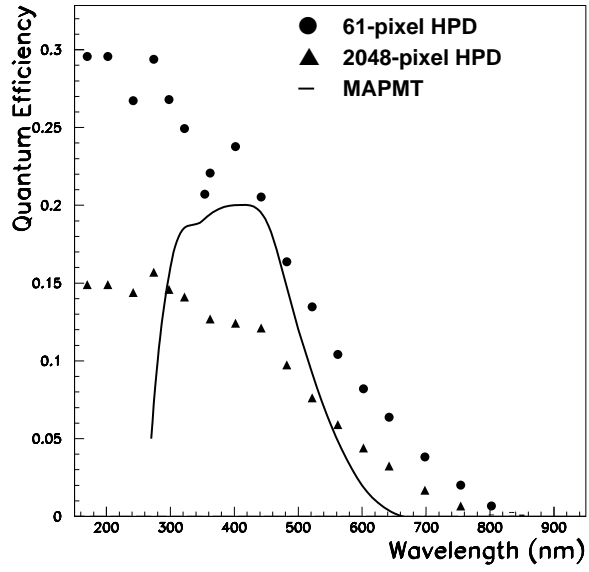,width=0.5\textwidth}
\end{center}
\caption{\label{qeff_para}
    \mbox{Quantum} Efficiency of the 2048-pixel HPD, a 
    61-pixel HPD and an MAPMT as a function of 
    the photon \mbox{wavelength}.}
\end{figure}

\begin{figure}
\begin{center}
\epsfig{file=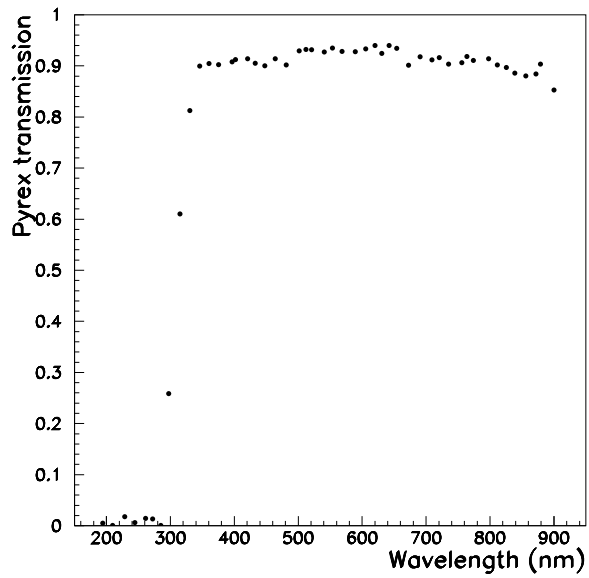,width=0.5\textwidth}
\end{center}
\caption{\label{pyrex_para}
 Measured transmission of the Pyrex filter 
 as a function of the photon \mbox{wavelength}.}
\end{figure}

\begin{description}
\item [Radiators:]{
 During different data-taking periods, air and 
$CF_{4}$ were used as radiators. The pressure and \mbox{temperature} 
of these radiators were monitored for correcting the 
refractive index \cite {t_paper2}. The gas circulation system 
which provided the $CF_{4}$ is described below.

During the $CF_{4}$ runs, data were taken at \mbox{various} pressures
ranging from 865 mbar to 1015 mbar and at different 
\mbox{temperatures} between $20^{0}C$ and $30^{0}C$. The
refractive index of $CF_{4}$ 
as a function of wavelength at STP using the 
parametrization in \cite {cf4_mea} is plotted in 
\mbox{Figure \ref {cf4_para}}.
}
\begin{figure}
\begin{center}
\epsfig{file=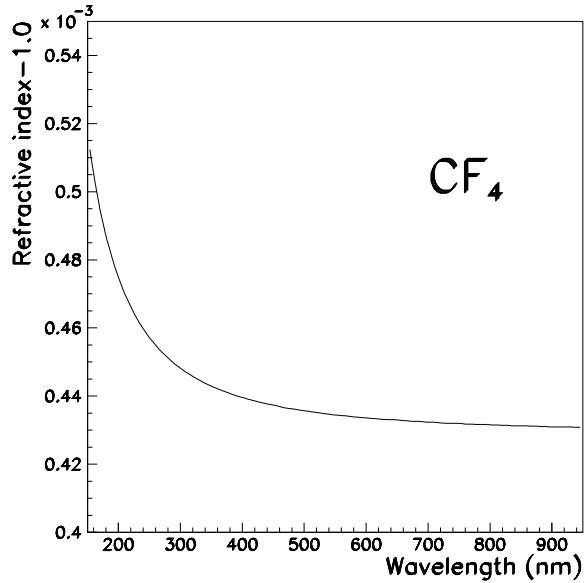,width=0.5\textwidth}
\end{center}
\caption{\label{cf4_para}
   Refractive index of $CF_{4}$ as a function of 
   the photon \mbox{wavelength} at STP using the
   parametrization in \cite{cf4_mea}.  }
\end{figure}
\item[$CF_{4}$ gas circulation system:]{As shown in the schematic 
diagram in Figure \ref{cf4circ}, the prototype Cherenkov vessel was 
connected into the gas circulation system, which was supplied 
by $CF_{4}$ gas \footnote { as supplied by CERN stores: 
reference SCEM 60.56.10.100.7} at high pressure. A molecular 
sieve (13X pore size) was included in the circuit to remove 
water vapour. The system used a microprocessor interface 
\footnote {Siemens S595U} to set and stabilise the required 
gas pressure and to monitor and record pressure, temperature 
and concentrations of water vapour and oxygen throughout the 
data taking. The absolute pressure of the $CF_{4}$ in the
Cherenkov vessel was maintained to within 1 mbar of the 
required value using electromagnetic valves which controlled 
the gas input flow and the output flow to the vent. Throughout 
the data taking the oxygen concentration was below 0.1$\%$ and 
the water vapour concentration was below 100 ppm by volume. }
\begin{figure}
\begin{center}
\epsfig{file=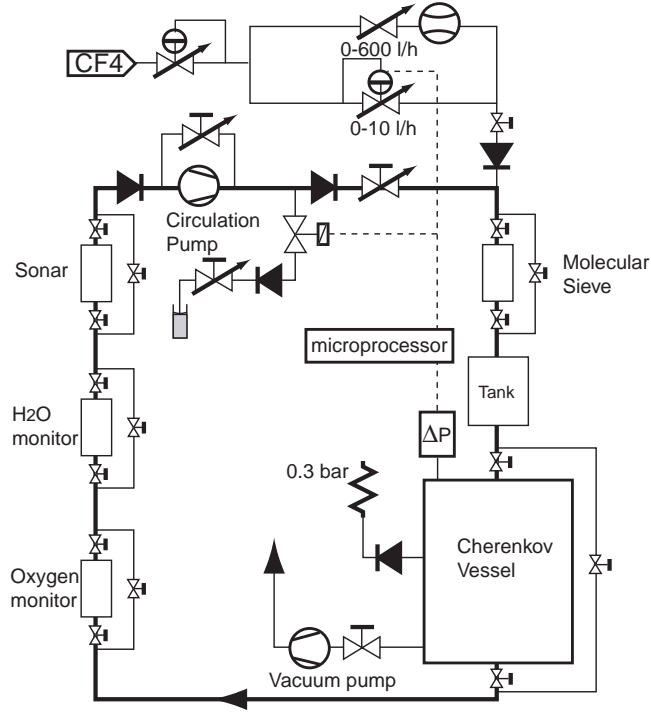,width=0.5\textwidth}
\end{center}
\caption{\label{cf4circ}
   Schematic diagram of the $CF_{4}$ gas circulation system.}
\end{figure}
\item[Mirror:]{
   The Cherenkov photons emitted were reflected by a mirror of
focal length 4003 mm which was tilted with respect to the beam axis
by 314 mrad, similar to the optical layout of the
LHCb RICH2. Using micrometer screws, the angle of tilt of the
mirror was adjusted to reflect photons on different regions 
of the photodetector plane which was located 4003 mm from the mirror.
The reflectivity of the mirror, measured as a function of the wavelength,
is shown in Figure \ref{mirror_para}.
}
\begin{figure}
\begin{center}
\epsfig{file=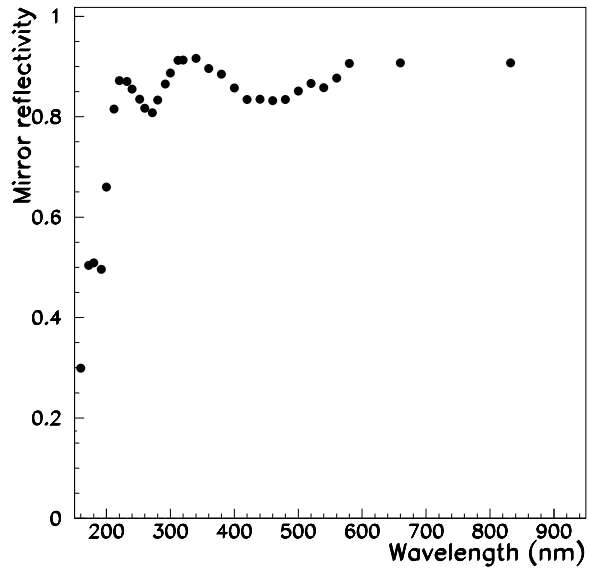,width=0.5\textwidth}
\end{center}
\caption{\label{mirror_para}Mirror reflectivity measured
    as a function of the photon \mbox{wavelength}.}
\end{figure}
\item[Photodetectors:]{
The important characteristics of the three different designs of
photodetectors tested are briefly summarised as follows:
{\begin{itemize}
\item[$\bullet$] {The 61-pixel Hybrid Photo-Diode (HPD) 
is manufactured by DEP and
has  an S20 (trialkali) photocathode deposited on a quartz window.
The quantum efficiency of a typical HPD measured by DEP, is plotted in
\mbox {Figure \ref {qeff_para}} as a function of the 
incoming photon wavelength.
Photoelectrons are accelerated 
through a 12 kV potential over 12 mm onto a 61-pixel silicon detector.
The image on the photocathode is magnified by 1.06 on the silicon
detector surface. This device gives an approximate gain of 3000. 
The pixels are hexagonally close packed and 
measure 2 mm  between their parallel edges. The signal is
read out  by a Viking VA2 \cite{viking} ASIC.} 
\item[$\bullet$] {The 2048-pixel HPD is manufactured in 
collaboration with DEP. 
It has \mbox{electrostatic} cross-focusing by which the image on the 
photocathode is demagnified by a factor of four at the 
anode. The \mbox{operating} voltage of this HPD is 20 kV. The anode has
a silicon detector, which provides an approximate gain of 5000, 
with an array of 2048 silicon pixels bump bonded to an LHC1 \cite{lhc1_chip} 
binary \mbox{readout} ASIC. Details of this device and its readout
can be found in \cite {thiery99}.

 Using the measurements made by DEP, the quantum efficiency 
of the S20 photocathode used on the 2048-pixel HPD is plotted 
in Figure \ref {qeff_para} as a function of the photon wavelength.
This tube has 
an active input window diameter of 40 mm and the silicon 
pixels are rectangles of size 0.05 mm $\times$ 0.5 mm. 
It represents a \mbox{half-scale} prototype of a final tube which 
will have an 80 mm diameter input window and 1024 square pixels 
with  0.5 mm side. }
\item[$\bullet$] {The 64-channel Multi-Anode PMT (MAPMT)
 is manufactured by Hamamatsu. It
has a bialkali photocathode deposited on 
a \mbox {borosilicate-glass} window 
and 64 square anodes mounted in an 8 $\times$ 8 
array with a pitch of 2.3 mm. The photoelectrons 
are \mbox{multiplied} using a 12-stage dynode chain resulting in an 
approximate overall gain of $10^{6}$ when operated at 900 V. 
From the measurements made by Hamamatsu, the quantum efficiency of a 
typical MAPMT as a function of the wavelength is shown 
in \mbox{Figure \ref{qeff_para}.}}
\end {itemize}}

During some runs, pyrex filters were placed in front of the 
photodetectors in order to limit the transmission 
to longer wavelengths
where the refractive index of the radiators 
is almost constant. In Figure \ref {pyrex_para} the transmission 
of pyrex as a function of photon wavelength is plotted.
}
\end {description}
\subsection{Experimental Configurations}
 The detector configurations used are summarised in Table \ref {table_photdet}.
   In configuration 1, seven 61-pixel HPDs and one MAPMT were
placed on a ring of radius 113 mm on the detector plate. In
configurations 2 and 3, a 2048-pixel HPD and three 61-pixel HPDs were placed
on a ring of radius 90 mm on the detector plate. In addition to these
configurations, the different radiator, beam and photodetector 
conditions used for the various runs are shown in Table \ref {table_config}.
\begin {table}[hbtp]
 \begin{center}
  \begin{tabular}{|c|c|c|}\hline
     Detector &   Detector    & Azimuthal \\ 
     Configu- &  Type & Location \\ 
     ration   &       & in Degrees \\ 
    \hline \hline
      1 & A1 & 310 \\ \cline {2-3}
                 & A2 & 345 \\ \cline {2-3}
     R=113 mm    & A3 & 75 \\ \cline {2-3}
                 & A4 & 105 \\ \cline {2-3}
     Radiator    & C & 150 \\ \cline {2-3}
     $CF_{4}$    & A5 & 195 \\ \cline {2-3}
                 & A6 & 230 \\ \cline {2-3}
                 & A7 & 270 \\ 
     \hline
       2 & B & 0 \\ \cline {2-3}
     R=90 mm            & A1 & 90 \\ \cline {2-3}
     Radiator     & A2 & 180 \\ \cline {2-3}
     Air             & A3 & 250 \\ 
     \hline
        3 & A3 & 340 \\ \cline{2-3}
     R=90 mm     & B   & 90 \\ \cline {2-3}
     Radiator    & A1 & 180 \\ \cline {2-3}
     Air         & A2 & 270 \\
     \hline
  \end{tabular}
  \caption{\label{table_photdet} Photon detector arrangement. 
   The zero degree \mbox{azimuthal} angle points 
   vertically up and increases anti-clockwise as seen from the mirror. 
   The symbols A1 to A7 denote 61-pixel HPDs, 
   B denotes the 2048-pixel HPD and C denotes the MAPMT. 
   R is the distance of each detector from the
   centre of the detector plate. }
  \end {center}
 \end {table}
 \begin {table} [hbtp]
 \begin {center}
   \begin{tabular}{|c|c|c|c|}\hline
     Run    & Beam     &  Detector  & Detectors \\
     Num- & Type     &  configu-  & with \\
     ber       & &  ration    & Pyrex \\
            &          &            & Filter \\
     \hline\hline
   1 & 120 GeV/{\it c} & 1                   & none \\ 
     &$ \pi^{-}$ &                     &     \\ \hline
   2 & 120 GeV/{\it c} & 1 & all HPDs \\ 
     & $\pi^{-}$ &                     &   \\ \hline
   3 & 100 GeV/{\it c}  & 2 & none \\ 
     & $\pi^{-}$ &                     &     \\  \hline
   4  &  100 GeV/{\it c} & 2 & all \\ 
     & $\pi^{-}$ &                     &     \\ \hline
   5 & 10.4 GeV/{\it c}  & 3 & none \\ 
    & $\pi^{-},e^{-}$ &                     &    \\ 
    &   &                 &    \\ \hline
   6 & 50 GeV/{\it c}&  1 & none \\
    & $\pi^{-},K^{-}$ &                     & \\ 
    &  &                      & \\ \hline    
  \end{tabular}
  \caption{\label{table_config} Experimental conditions for
   the various runs.}
 \end{center}
\end {table}
\subsection{Data Acquisition System} 
The 61-pixel HPDs and the MAPMT use analogue readout 
whereas the \mbox{2048-pixel} HPD uses binary readout. 
A detailed description of their respective data acquisition 
systems can be found in \cite {t_paper1} and \cite {thiery99}.

For the analogue readout system, the mean and
width of the pedestal distributions for each pixel were 
calculated using dedicated pedestal
runs, interleaved between data runs triggered with beam. 
Some data were also taken using light emitted from a pulsed 
Light Emitting Diode (LED) for detailed studies of the 
photoelectron spectra. Zero suppression was not used on analogue data 
from the photodetectors.

\begin{figure}
\begin{center}
\epsfig{file=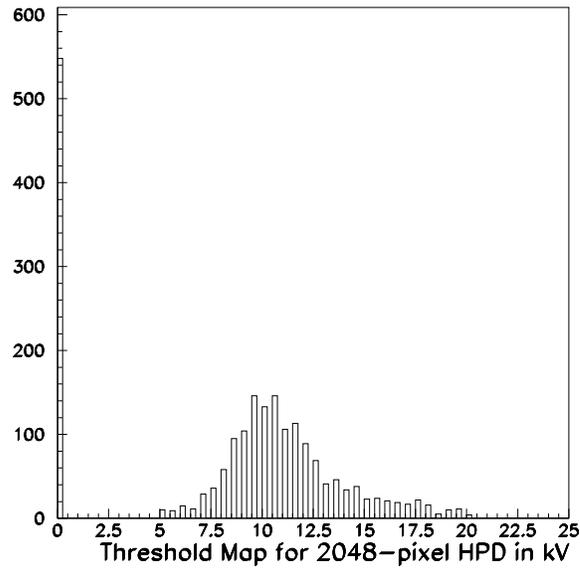,width=0.5\textwidth}
\end{center}
\caption{\label{thrmap} Threshold map for the binary readout
   system of 2048-pixel HPD. \mbox{Thresholds} shown are in kV. 
   The pixels indicated as having zero threshold (first bin) 
   are those with too low or too high a threshold.}
\end{figure}

A pixel threshold map was established on the 2048-pixel HPD 
using an LED \cite {thiery99}. For this, the high voltage applied
on the tube was varied, and the voltage for each channel to become 
active was recorded. This
threshold map was used to identify pixels 
with too low a threshold, which were then masked. It was also used to
identify pixels with too high a threshold and hence insensitive to
photoelectrons. A histogram of the threshold map is shown in 
Figure \ref {thrmap} where the pixels which were masked or 
insensitive (26$\%$) are indicated by the entries in the first bin.
For this device, the noise ($\sigma_{N}$) of the readout electronics 
is 160 electrons (0.6 kV Silicon equivalent) and the distribution 
of the silicon pixel thresholds has an rms width of \mbox{1.6 kV}. 

\begin{figure}
\begin{center}
\includegraphics[width=0.5\textwidth]{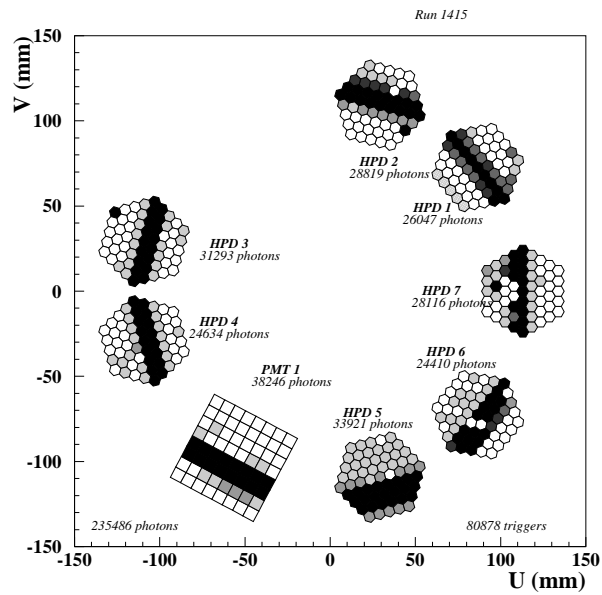}
\end{center}
\caption{\label{ev1disp} Event Display for a run in 
configuration 1. For clarity the photodetectors are
 magnified.}
\end{figure}

In Figure \ref {ev1disp} an online display, integrating 
all events in a run, with seven 61-pixel HPDs and an MAPMT 
in configuration 1 is shown. 
Part of the Cherenkov ring falls on the \mbox{photodetectors} 
and is clearly visible.
\section {\large Simulation of RICH2 prototype}
The RICH2 prototype configurations are simulated to allow detailed 
\mbox{comparisons} of expected performance with that found in data. 
The simulation program generates photons uniformly in energy and with
the corresponding Cherenkov angle. The trajectories
of these photons, and the photoelectrons they produce, are simulated 
using the beam divergence, beam composition and 
the optical characteristics of the various components of the 
RICH detector shown in Figures \ref {cf4_para} to \ref {pyrex_para}.
The air radiator is simulated using a 
gas mixture consisting of 80\% Nitrogen and 20\% Oxygen.

The program also simulates the response of the various photodetectors.
Since the 2048-pixel HPD used binary readout, to study its response
the program simulates the threshold map ( Figure \ref {thrmap}) 
used for this readout. The simulation of the response
of the silicon detector of this HPD is described in Section 4.1

\section{\large Estimates of Photoelectron yield}
 The average number of photoelectrons detected per event in a photodetector 
defines the photoelectron yield for that detector. This 
is determined for the configurations 1 and 2 indicated in 
Table \ref {table_photdet}. 
Since the 61-pixel HPD and the MAPMT use analogue readout, the 
distinction between signal and background depends upon the 
threshold above the pedestal peak assigned to the measured 
photoelectron spectrum. To get the true photoelectron yield 
at a given threshold, estimates are made for the level of 
background present and for the amount of signal loss 
that occurs as a result of applying the threshold cut, specified in
terms of the width ($\sigma$) of the pedestal spectrum. 

In the two types of HPDs, there is 18$\%$ probability \cite {bsca_ref}
at normal incidence, for electrons to backscatter at the silicon surface, 
causing some loss of signal. In the 61-pixel HPD, the backscattered 
electrons can ``bounce`` off the silicon surface more than once, 
whereas in the 2048-pixel HPD the electric field is such that they do not
return to the silicon detector. 
Passage through the dead layers of the silicon wafer can also 
cause a small amount of signal loss in the HPDs. Since the 2048-pixel HPD 
uses binary readout, its photoelectron yield depends mainly upon 
the threshold map of the readout system. 

From the estimate of the photoelectron yield ($N_{pe}$) of a photodetector,
the figure of merit ($N_{0}$) is calculated using: \\
\begin{math}
  N_{0}=N_{pe}/(\epsilon_{A} L \sin^{2}\theta_{c}) \\
\end{math}
 where $\epsilon_{A}$ is the fraction of the Cherenkov ring covered by the
photodetector, L is the length of the radiator and $\theta_{c}$ is the
mean Cherenkov angle measured using the method described in \mbox{Section 5.}
\subsection{Photoelectron yield for the 2048-pixel HPD}

The response of the silicon detector of 
this HPD is simulated as follows: 

Each photoelectron is accelerated through a 
potential of 20 kV towards the silicon surface. 
The probability for backscattering at the silicon surface is
18 $\%$\cite {bsca_ref}. During the backscattering process, 
only a fraction of the 20 keV energy is released in the silicon detector. 
For an energy release varying from 5 to 20 keV, the energy loss in the
dead layer of the silicon ranges from 5 to 1.2 keV as described 
in \cite {thiery99} and references therein. 
A readout channel is expected to fire only when the charge signal 
generated in the silicon detector exceeds
the corresponding pixel threshold by at least 4 times the electronic noise.
%
%\begin{figure}
%\begin{center}
%\epsfig{file=sigloss.eps,width=0.6\textwidth}
%\end{center}
%\caption{\label{sigloss} Signal Loss (in kV) vs. applied signal (in kV)
%    for the Silicon Detector in 2048-pixel HPD.}
%\end{figure}
%

\begin{figure}
\begin{center}
\epsfig{file=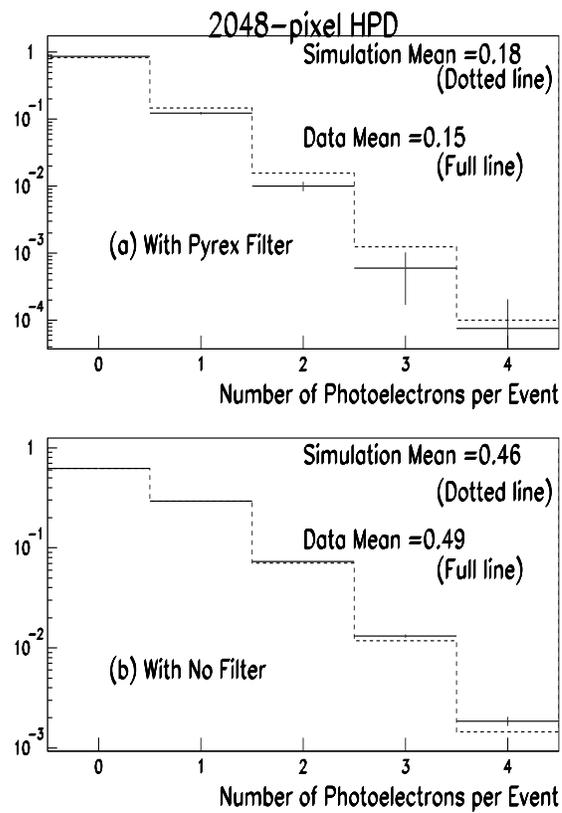,width=0.5\textwidth,height=0.7\textwidth}
\end{center}
\caption{\label{ispamult} Number of photoelectrons per event 
  in the 2048-pixel HPD 
  (a) with pyrex filter in simulation and real data
  (b) with no filter in simulation and real data.}
\end{figure}

 A flat background of 0.01 photoelectrons per event is observed 
in the real data on the detector surface from beam related
sources such as photons and photoelectrons reflected in random directions from 
different surfaces in the prototype. This 
is also incorporated into the simulation. The resultant 
\mbox{photoelectron} yield from the simulation in the
presence of a pyrex filter is shown in Figure \ref {ispamult}(a), 
and in the absence of any filter is shown in Figure \ref {ispamult}(b).

The systematic error in the photoelectron yield 
is evaluated from the simulation by varying the parameters
which are listed below. The result of these variations are tabulated in
\mbox{Table \ref {ispaphy}}.
{\begin{itemize}
\item[$\bullet$]{Quantum efficiency of the phototube: The quantum
efficiency of the 2048-pixel HPD is found to be approximately half that
of the 61-pixel HPD. The simulation is repeated by replacing 
the quantum efficiency of the 2048-pixel HPD
 with those from the 61-pixel HPD, scaled down by a factor
 of two.}
\item[$\bullet$]{Amount of photon absorption in oxygen: The simulation
 is repeated with and without activating the photon absorption although
 this is significant only for wavelengths below 195 nm.}
\item[$\bullet$] {Wavelength cutoff of the photocathode: To account for
any variation in the active wavelength range among different versions of
the photocathodes, the simulation is repeated by varying lower cutoff
between 190 nm and 200 nm, and the upper cutoff between 600 nm and 900 nm.} 
\item[$\bullet$] {Backscattering probability at the silicon surface: The
simulation is repeated by varying the backscattering probability between
16$\%$ and 20 $\%$.}
\end {itemize}}
\begin {table} [hbtp]
 \begin {center}
   \begin{tabular}{|c|c|c| } \hline
              &\multicolumn{2}{|c|}{Variation of yield}  \\ \hline  
    Simulation  & with  & with no  \\ 
    parameters           & Pyrex   & Filter \\ \hline
    Quantum &0.020 &0.030\\ 
    Efficiency  & &      \\ \hline
    Oxygen  & 0.003 &0.060 \\ 
    Absorption  & &  \\ \hline
    Active &0.003 & 0.030 \\ 
    Wavelength & & \\
    Range & & \\ \hline
    Back-  & 0.006 & 0.014 \\ 
    scattering  &  &  \\  \hline
    Overall & 0.021 & 0.074 \\ 
    Variation & & \\ \hline
   \end {tabular}
  \caption {\label{ispaphy} Variation of the simulated photoelectron yield
   in the 2048-pixel HPD under different conditions. The amount of
   variation of each of the simulation parameters is described in the
   text.}
  \end {center}
\end {table}
The simulated photoelectron yield per detector
in the case without any filter 
is 0.46 $\pm$ 0.07, whereas in real
data the yield is 0.49 (Figure \ref {ispamult} (b)). 
The simulated yield per detector, for the case with the pyrex filter, 
is 0.18 $\pm$ 0.02 and the corresponding yield in real data 
is 0.15 (Figure \ref {ispamult} (a)). Using these yields, the figure of
merit is estimated to be \mbox{97 $\pm$ 16 $cm^{-1}$} 
in the case without any filter and \mbox{30 $\pm$ 5 $cm^{-1}$} 
in the case with the pyrex filter. For the case without any filter,
an independent determination \cite {comoproc} 
of the figure of merit for the same tube, agrees with the present estimate.
\subsection{Photoelectron yield for the 61-pixel HPD}

\begin{figure}
\begin{center}
\epsfig{file=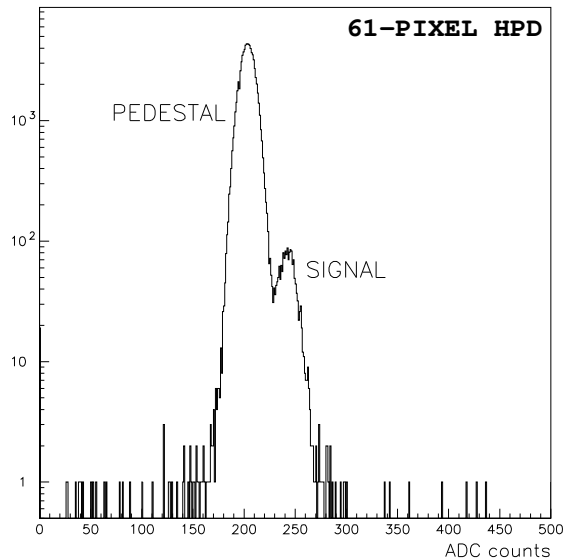,width=0.5\textwidth}
\end{center}
\caption{\label{adchpd} ADC distribution of Cherenkov photoelectrons
 in 61-pixel HPD.}
\end{figure}

 Figure \ref {adchpd} shows a typical photoelectron spectrum obtained
from a single pixel in a 61-pixel HPD. The
peaks corresponding to the pedestal and signal can be clearly seen.
In similar distributions obtained for each of the pixels, the 
background contamination in the photoelectron yield 
and the amount of signal lost are estimated as
a function of the threshold cut using two different analysis methods. 
One of these methods is described below and the other one is described
in Section 4.3 where similar estimates are made for the MAPMT.
\begin{description}
\item [Signal loss estimate:]{
 The signal loss is estimated using data where the
signals were provided by photons from an LED as only these runs
have adequate statistics for this purpose. The signal
loss is considered to have a Gaussian component and a backscattering component
which are described below.

An example of the spectra for each detector pixel in LED data is shown in
Figure \ref {adcled}. It can be divided mainly into three parts 
identified as \mbox{distributions} for 
the pedestal, one photoelectron and two photoelectrons, in addition to
two underlying distributions corresponding to the backscattering 
contributions to the single and double photoelectron spectra. 
In order to estimate these backscattering contributions, a 
backscattering probability of 18$\%$ \cite {bsca_ref} 
is \mbox{assumed.} The energy distribution of the backscattered 
electrons is made by convoluting the distribution of the energy fraction
of the backscattered electrons for \mbox{10 keV} electrons 
incident on aluminium, obtained from \cite {bsca_ref}, with a Gaussian
that has the same width as that of the pedestal spectrum in LED data.

The adc spectrum in LED data is fitted with a function that modelled
the spectrum as a sum of three Gaussians
with contributions from two backscattering components.
The three Gaussians correspond to the
distributions of pedestal, one photoelectron and two photoelectrons.
The result of the fit is superimposed over the adc spectrum 
in Figure \ref {adcled}. The widths of the Gaussians for the 
photoelectrons are then corrected to account for the slight difference 
in the widths of the pedestal observed in LED data and Cherenkov photon data. 

In the region below the threshold cut, the sum of the area which is
under the one photoelectron Gaussian and the corresponding 
backscattering component is then taken as the sum of the 
Gaussian and backscattering components of the signal loss.

This procedure is repeated using a different LED run and varying the
backscattering probability between 16$\%$ and
20$\%$. The resultant variations obtained in the signal loss estimate 
are taken as contributions to systematic error from this method.

At the threshold cut of 3$\sigma$, the Gaussian component of 
the signal loss is 0.9$\%$ whereas the backscattering component is 11.2$\%$.

\begin{figure}
\begin{center}
\epsfig{file=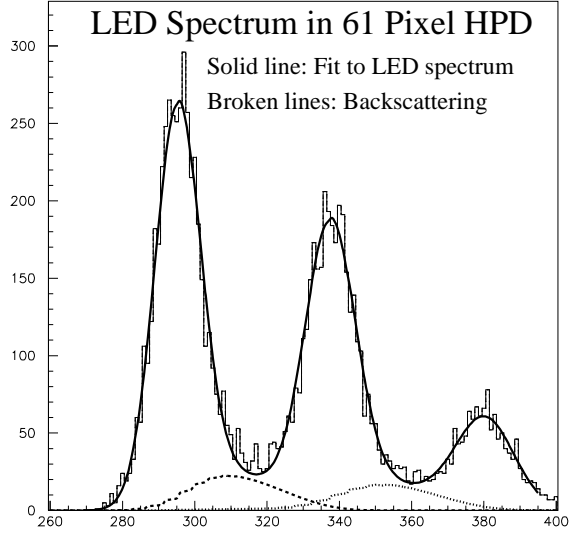,width=0.5\textwidth}
\end{center}
\caption{\label{adcled} LED spectrum in 61-pixel HPD. The three peaks
correspond to 0, 1 and 2 photoelectrons. Backscattering contributions 
estimated from \cite {bsca_ref}  are shown as broken lines.}
\end{figure}
}

\item[Background estimate:]{ 

 The background remaining in the Cherenkov photoelectron spectrum after 
a given threshold cut is considered to have a Gaussian component due 
to electronic noise, and a non-Gaussian component induced by detector 
noise and photons from extraneous sources. For the first component, 
a single Gaussian is fit to the pedestal part of this spectrum. 
The area under this fit spectrum above the threshold cut is then 
taken as the Gaussian component of the background. 
This procedure was repeated changing the upper range of the Gaussian fit
from 1.2$\sigma$ to 2$\sigma$ and the resultant variation in the
background estimate is taken as a contribution to the systematic error.

 In order to evaluate the second component, data from pedestal runs are used.
The fraction of the spectrum above the threshold cut, after removing 
the fit single Gaussian to the pedestal spectrum, 
is taken as the non-Gaussian component. The variation in this
estimate obtained using different pedestal runs is taken as a contribution to
the systematic error.

After correcting the distribution of the number of 
photoelectrons in each pixel for background 
and signal loss, their spatial distribution on the silicon surface is fitted 
with a function which assumes the Cherenkov angle distribution 
to be a Gaussian. A residual flat background observed in this fit 
is considered as beam related background and is subtracted from the
photoelectron signal. The fit is repeated by varying the 
parameters of the function and the resultant variations in the 
background estimate is taken as a contribution to the systematic error.
}
\item [Estimates of yield:]{

The results obtained for the photoelectron multiplicities
 after correcting for background and signal loss using the above
method are reported below. These are in agreement with the 
results obtained from the alternative method described in the next section.

In these estimates, the statistical error 
is found to be negligible compared to the overall
systematic error which is obtained by adding the various
contributions in quadrature. The contributions to
the systematic error are shown in Table \ref {hpd_ysys}.
In Table \ref {hpdyield}, the corrected photoelectron yields
for the data with pyrex filter and
with no filter are shown along with the corresponding 
expectations from simulation. The yields from data and simulation
agree.

As a systematic check, the stability of the corrected 
photoelectron yields obtained by varying the threshold cut 
from 2$\sigma$ to 5$\sigma$ for the 
data with pyrex filter, is shown in Table \ref {hpdsigma}.
The small variation seen in the yields between 3$\sigma$ and 4$\sigma$
is quoted as a systematic error contribution in Table \ref {hpd_ysys}. 
The fact that the corrected photoelectron yields estimated are 
independent of the threshold cut and that the two analysis methods yield 
similar results give confidence in the results shown in Table \ref {hpdyield}.

Using the yield estimates in Table \ref{hpdyield}, the figure of merit is
estimated to be \mbox{89 $\pm$ 8 $cm^{-1}$} in the case with pyrex filter 
and \mbox{258 $\pm$ 24 $cm^{-1}$} in the case without any filter.

}
\end{description}
%
%\begin {table} [hbtp]
% \begin {center}
%   \begin {tabular} {|c|c|c|} \hline
%          Analysis     & Method 1 & Method 2 \\  \hline
%     Background        & 0.08      & 0.07       \\  \hline
%     Signal            & 18.2      & 14.2       \\ 
%     Loss \%           &           &            \\ \hline
%     Photo-            & 0.29$\pm$ 0.03 & 0.28$\pm$ 0.02  \\ 
%     electron          &          &            \\
%     yield             &          &            \\ 
%                       &          &            \\
%                       &          &            \\  \hline
%   \end {tabular}
%  \caption {\label{yieldcom} Comparison of the average 
%              photoelectron yields after corrections between
%              two analysis methods for 61-pixel HPD 
%              in configuration 1 
%               using pyrex filter. 
%              Results above a threshold of 3 $\sigma$ are used.}
%  \end {center}
%\end {table} 
%
\begin {table} [hbtp]
 \begin {center}
   \begin {tabular} {|c|c|c|} \hline
   Error Source  & with & No \\ 
                 & pyrex & Filter\\ \hline \hline
   Background :  &    &  \\ 
                 &    &   \\
                 &    &   \\
   Electronics   & 0.003 & 0.011 \\
   Noise         &       &        \\
   (b)Detector   & 0.005 & 0.009 \\
    Noise        &       &       \\
   (c)Beam       & 0.001 &0.010 \\ 
   Related       &     &      \\ \hline 
   Signal loss:  &     & \\
   (a)backscattering & 0.004   & 0.012 \\
     fraction        &         &        \\
   (b)Fit Stability & 0.004 &0.014 \\ \hline
   Change of & 0.002 & 0.007 \\ 
   Threshold &       &       \\
\hline \hline
   Overall error & 0.008  & 0.027 \\ \hline
    
   \end {tabular}
  \caption {\label{hpd_ysys} Contributions to the
              systematic error for the photoelectron yield
              estimate, in 61-pixel HPD.
              Errors from background and signal loss estimates
              shown are at 3 $\sigma$ threshold. } 
  \end {center}
\end {table} 
\begin {table} [hbtp]
 \begin {center}
   \begin {tabular} {|c|c|c|c|} \hline
     Photo-    & Using & Using \\
     electron  & Pyrex & No     \\
     yield     & Filter & Filter  \\
               &        &         \\
               &      &         \\
\hline
Real   &  0.29 $\pm$ 0.01  & 0.86 $\pm$ 0.03 \\ 
Data   &                    &                  \\    \hline
Simulated &  0.31       &    0.86   \\
Data   &                &            \\ \hline
   \end {tabular}
   \caption {\label{hpdyield} The average photoelectron yields 
    per detector after corrections,
    above a \mbox{3 $\sigma$} threshold for 61-pixel HPD in
    \mbox{configuration 1} in real and simulated data 
    using $CF_{4}$ radiator.}
 \end {center}
\end {table} 
\begin {table} [hbtp]
 \begin {center}
   \begin {tabular} {|c|c|c|c|} \hline
 Threshold   & Back-   & Signal  & Photo- \\
 in          & ground  & Loss   &   electron  \\ 
 Number      &         & (\%)     &  yield  \\ 
 of $\sigma$ &         &   &   \\ \hline 
        2    & 1.36 & 7.8 &  0.29 \\ \hline
        3    & 0.19 &  12.1 &  0.29 \\ \hline
        4    & 0.08 & 18.2 &  0.29 \\ \hline
        5    & 0.06 & 31.4 &  0.30 \\ \hline
   \end {tabular}
 \caption {\label{hpdsigma} The average photoelectron yield per
   detector after corrections
  in 61-pixel HPD in configuration 1 for
  different values of the \mbox{threshold}. Estimates of the background and
   signal loss are also shown.}
 \end {center}
\end {table} 
\subsection{\large Photoelectron yield for MAPMT}
\begin{figure}
\begin{center}
\epsfig{file=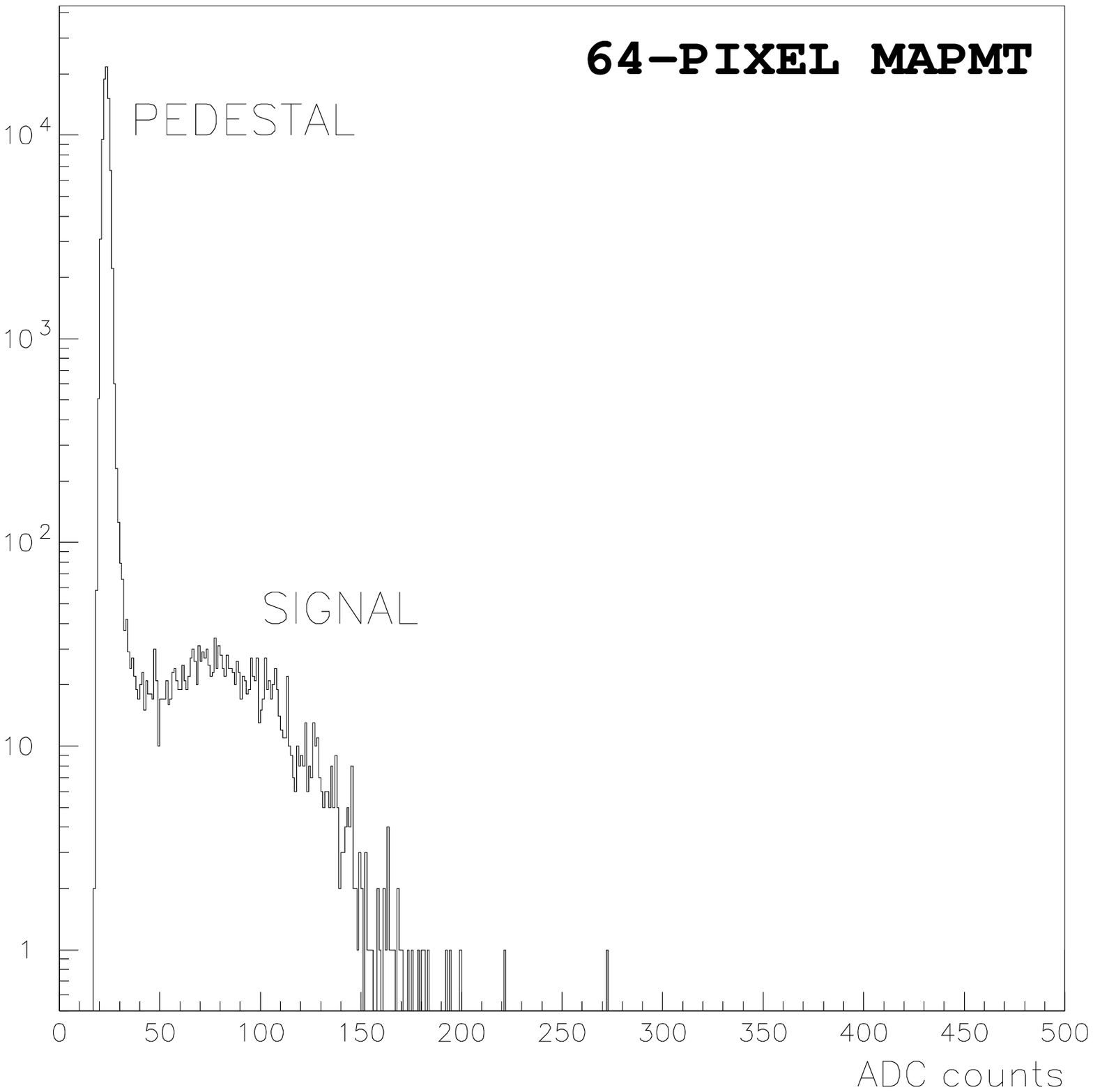,width=0.5\textwidth}
\end{center}
\caption{\label{adcpmt} ADC distribution of Cherenkov photoelectrons
 in MAPMT.}
\end{figure}
 Figure \ref {adcpmt} shows a typical pulse height distribution for
a pixel in the MAPMT in beam triggered runs. The photoelectron signal and 
pedestal peaks can be clearly distinguished. The amount of 
signal lost and the amount of background contamination 
to the photoelectron yield are estimated using the method described below.

\begin{description}
\item [Signal loss estimate:]{
 This method also uses data where the photons from an LED 
provided signals to the MAPMT. A Gaussian is fit to the pedestal part
of the pulse height distribution. The contribution of the pedestal
is removed, and in the remaining spectrum that part
below the threshold cut is taken to be the signal loss.
The contributions to the systematic
error in this estimate are listed below:
\begin{itemize} 
\item[$\bullet$]{The change in signal loss
 obtained by swapping the width of the pedestal in Cherenkov
photon data with that from LED data, is taken as a contribution to the
systematic error.}
\item[$\bullet$]{ In the Cherenkov photon data and LED data, 
the ranges of the fits to the pedestals are varied and  
any resultant change in the signal loss is taken as the
contribution to the systematic error. }
\end {itemize}
}

\item [Background estimate:]{
  In order to estimate the background level, data from a
special run are used where the pressure in the $CF_{4}$ radiator was reduced 
such that the Cherenkov ring passed through a different set of pixels
than in the other runs. In these data, the photoelectron yield 
is estimated after applying the threshold cut to the spectrum 
from the pixels which are selected to be off the Cherenkov ring. 
Assuming a uniform background across the MAPMT, this yield is taken 
as the background contribution. This procedure is repeated by 
varying the set of pixels which are selected for this estimate and 
the resultant change in the background estimate is taken as  
contribution to the systematic error.
}
\item [Estimate of yield:]{

These estimates for the background level and signal loss are repeated
for different threshold cuts in the spectra with the results given
in Table \ref {pmtyield}. The photoelectron yields resulting from 
these estimates are independent of the threshold cuts applied.
The systematic error in this measurement is estimated in the same way as 
for the 61-pixel HPD described in the previous section.
Above a threshold cut of 3 $\sigma$, the yield after the corrections 
is estimated to be 0.48 $\pm$ 0.03. The
corresponding expectation from simulation is 0.52. The discrepancy
between data and simulation is attributed to the uncertainty in the 
knowledge of the quantum efficiency of the particular MAPMT used in 
these tests. Using this yield estimate, the figure of merit 
is estimated to be \mbox{155 $\pm$ 13 $cm^{-1}$}.
}
\end {description}
\begin {table} [hbtp]
 \begin {center}
   \begin {tabular} {|c|c|c|c|} \hline
     Thre-       & Back- & Signal & Number   \\
     shold in    & ground & Loss & of Photo-  \\
     Number      &  & (\%)     & electrons \\
     of $\sigma$ &  &  & per Event \\
 \hline
    2  & 1.87 & 0.4 & 0.43 $\pm$ 0.07 \\ \hline
    3  & 0.17 & 3.0 & 0.48 $\pm$ 0.03  \\ \hline
    4  & 0.02 & 6.0 & 0.47 $\pm$ 0.03 \\ \hline
    5  & 0.01 & 8.5 & 0.47 $\pm$ 0.03 \\ \hline 
%   \hline
%   Simulation & 0.44 & & \\ \hline
   \end {tabular}
  \caption {\label{pmtyield} The photoelectron yield in the MAPMT after
  the corrections, background estimates and signal loss estimates for
  different threshold cuts on the ADC spectra.}
 \end {center}
\end {table} 
%
%\clearpage
\section{\large Resolution of the Reconstructed Cherenkov Angle}

   As described in \cite {t_paper2}, the reconstruction of the
Cherenkov angle requires the coordinates of the hit on the photodetector,
the centre of curvature of the mirror and the photon emission point (E)
which is assumed to be the middle point of the track in the radiator. 
The point (M) where the photons are reflected off the mirror, 
is reconstructed using the fact that it lies in the plane defined by
the aforementioned three points. The reconstructed Cherenkov angle is the 
angle between the beam direction and the line joining E and M.
  
\begin{figure}
\begin{center}
 \epsfig{file=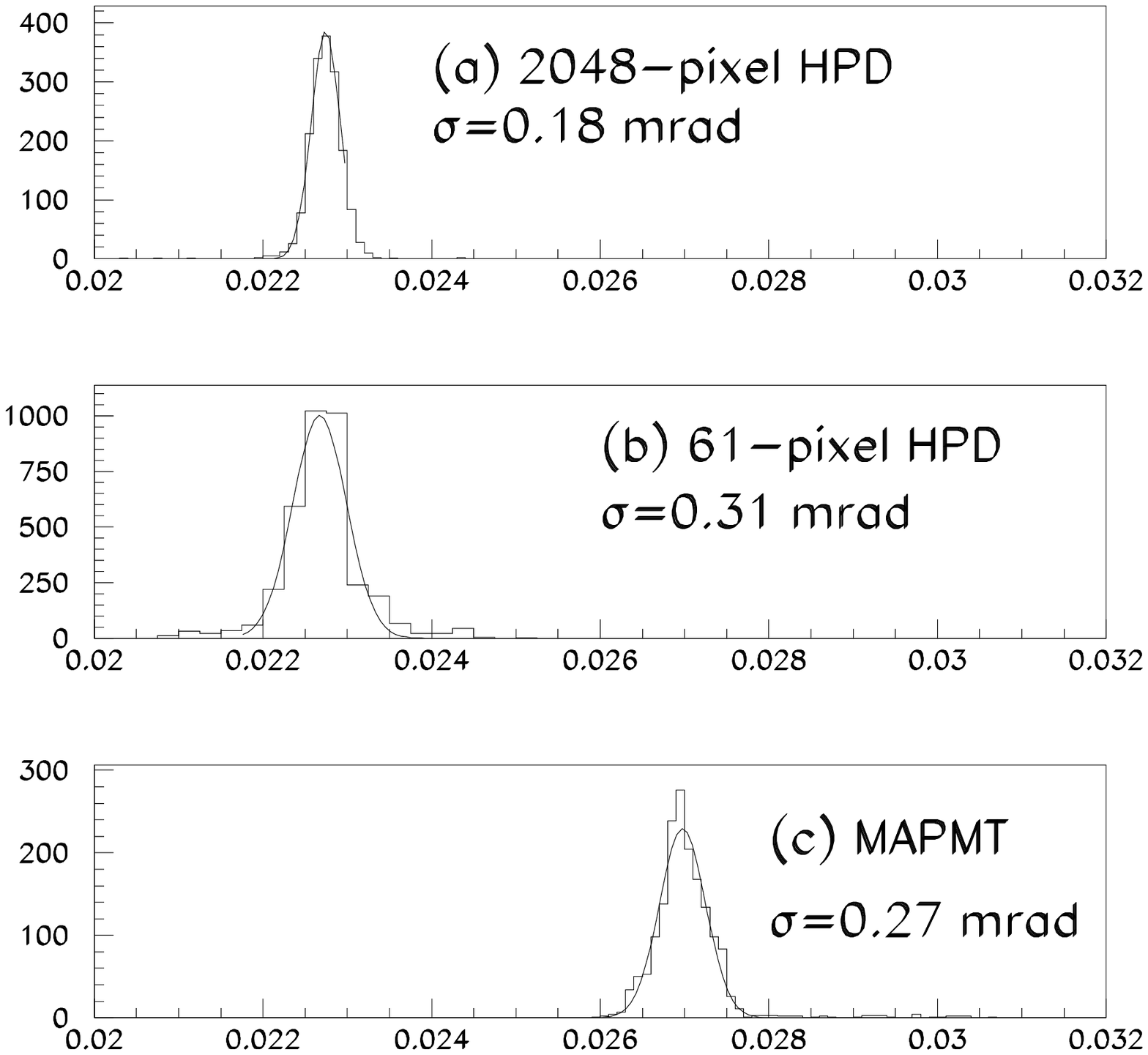, width=0.5\textwidth,height=1.2\textwidth}
\end{center}
\caption{\label{ckvang} Cherenkov Angle Distribution for (a) 2048-pixel
 HPD (b) 61-pixel HPD (c) MAPMT. The two HPDs were in 
 configuration 2 with
 pyrex filter and were diametrically opposite to each other on the 
 Cherenkov ring. The  MAPMT was in 
 configuration 1. 
 The various configurations are listed in Table \ref {table_photdet}. }
\end{figure}

  Figures \ref {ckvang}(a),(b) show the Cherenkov angle distribution 
  obtained using air radiator and 100 GeV/{\it c} pions 
  for the 2048-pixel HPD and a 61-pixel HPD which were diametrically 
  opposite to each other on the detector plate in 
  configuration 2 with pyrex filter. The 2048-pixel HPD has a 
  better resolution 
  than the 61-pixel HPD since the pixel granularity is 0.2 mm for the 
  former and 2 mm for the latter. Figure \ref {ckvang}(c) shows the 
  Cherenkov angle distribution obtained using $CF_{4}$ radiator and
  120 GeV/{\it c} pions for an MAPMT with 2.3 mm pixel granularity 
  in configuration 1. 
\subsection{Sources of Uncertainty in the Cherenkov Angle Measurement}
 {\begin{itemize}
 \item[$\bullet$]{ Chromatic Error: This is due to the variation of refractive
      index of the radiator with wavelength and is largest in the
      UV region. Use of pyrex filters reduces this contribution.}
 \item[$\bullet$]{Emission point uncertainty: This comes from the 
      fact that the
      mirror is tilted with respect to the beam axis and that the emission
      point is assumed to be in the middle of the radiator, regardless 
      of the true but unknown point of emission.}
 \item[$\bullet$]{Pixel size of Photodetector.}
 \item[$\bullet$]{Measurement of beam trajectory: This contribution 
       comes from the granularity of the pixels in the silicon detectors 
       which are used to measure the direction of the incident beam particle.}
 \item[$\bullet$]{Alignment: This contribution comes from residual 
       misalignments between the silicon telescope, the mirror and the 
       photodetectors.}
 \end {itemize}}
     In Table \ref {tableres} the resolutions from each of the
   above components are tabulated for each of the three photodetectors
   in typical configurations. In each case, the overall simulated 
   resolution is in good agreement with that measured in the 
   beam triggered data.

\begin {table} [hbtp]
 \begin {center}
   \begin {tabular} {|c|c|c|c| } \hline
    Resolution & 2048- & 61- & MAPMT \\ 
    Component     & pixel&  pixel     &       \\ 
                & HPD &HPD  & \\ \hline
    Chromatic  & 0.15 & 0.13 & 0.14 \\
    Error  & & & \\ 
    (with Pyrex    &  &  &   \\
     on HPDS)      & & &  \\ \hline 
    Emission  & 0.05 & 0.05 & 0.08 \\ 
    Point & & & \\ \hline
    Pixel     & 0.02  & 0.13 & 0.17 \\ 
    Size & & & \\ \hline
    Telescope  & 0.06 & 0.06 & 0.06 \\ 
    Pixel & & & \\
    Size  & & & \\ \hline
    Alignment & 0.06 & 0.08 & 0.10 \\  \hline \hline
    Overall MC & 0.17 & 0.21 & 0.26 \\ \hline
    Overall Data & 0.18 & 0.26 & 0.27 \\ \hline
   \end {tabular}
   \caption {\label{tableres} Resolution components 
    in mrad for the three photodetectors. 
    The 2048-pixel HPD was in 
    configuration 2 and the other two detectors were 
    in configuration 1. The second 61-pixel HPD from 
    configuration 1 is used in this table. }
 \end {center}
\end {table}

\begin{figure}
\begin{center}
\epsfig{file=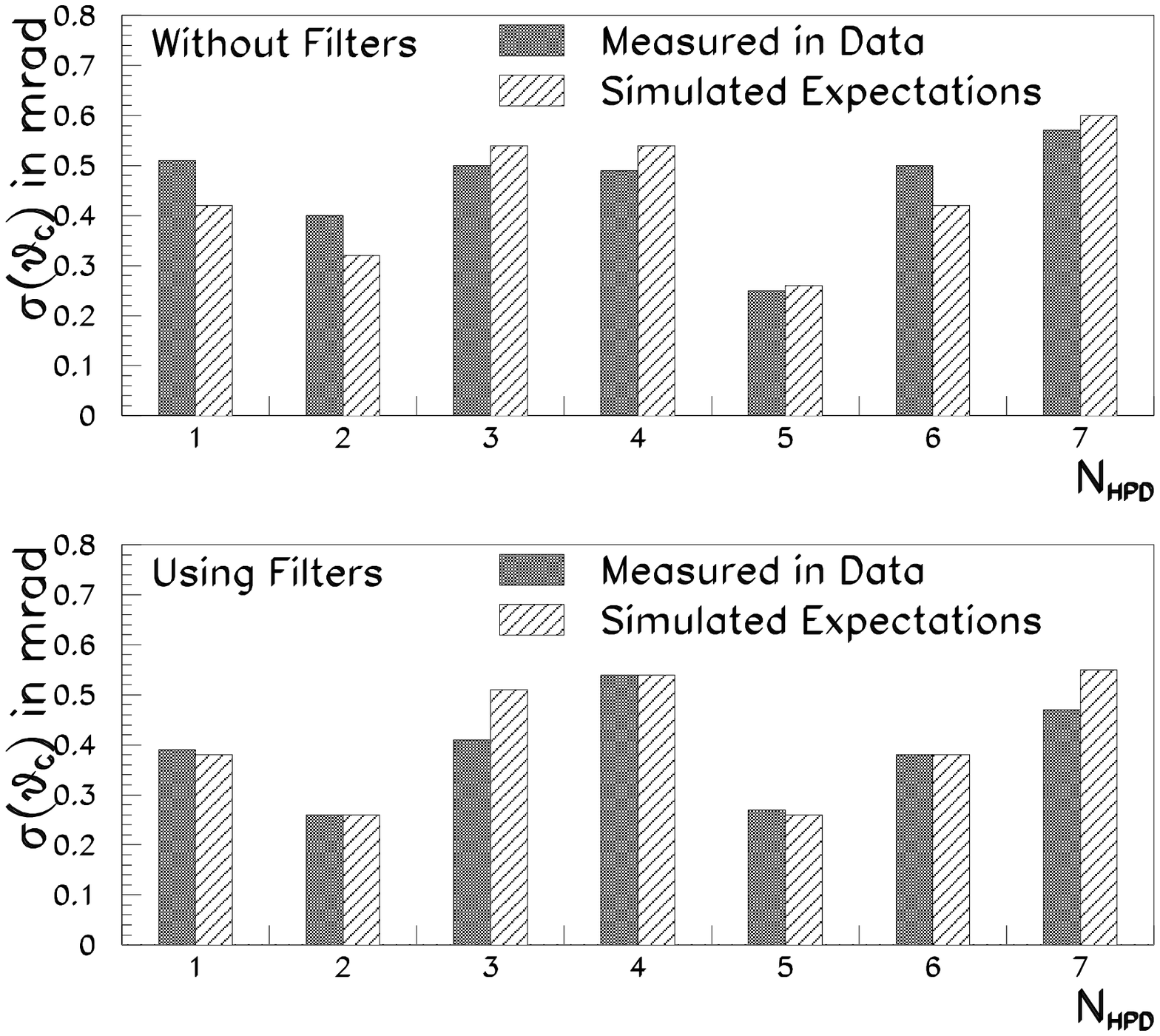,width=0.5\textwidth}
\end{center}
\caption{\label{hpdres} Cherenkov Angle Resolutions in mrad 
   for seven 61-pixel HPDs in configuration 1
  (a) without pyrex filter and (b) with pyrex filter.}
\end{figure}

    In configuration 1 with seven HPDs
    it was possible to perform a detailed investigation of the
    Cherenkov angle resolution. Figure \ref {hpdres}(a) shows 
    the resolution measured in data and from simulation for each 
    of the seven 61-pixel HPDs in this configuration.
    Agreement is seen between data and \mbox{simulation} in all 
    cases. Each HPD in this figure was located at a different azimuth
    on the detector plate and hence has a different
    emission point uncertainty. Hence the overall resolution for
    different HPDs are different. Figure \ref {hpdres}(b) shows 
    the same resolutions, for the data using the pyrex filter, 
    which reduces the contribution from chromatic error.

   The expectation from the LHCb Technical proposal 
    \cite {lhcb_proposal} is to have a resolution of 0.35 mrad 
    which is already achieved for the MAPMT, the 2048-pixel HPD 
    and some of the HPDs shown in Figure \ref {hpdres}.

   \subsection{Multiphoton Resolution}
\begin{figure}
\begin{center}
\epsfig{file=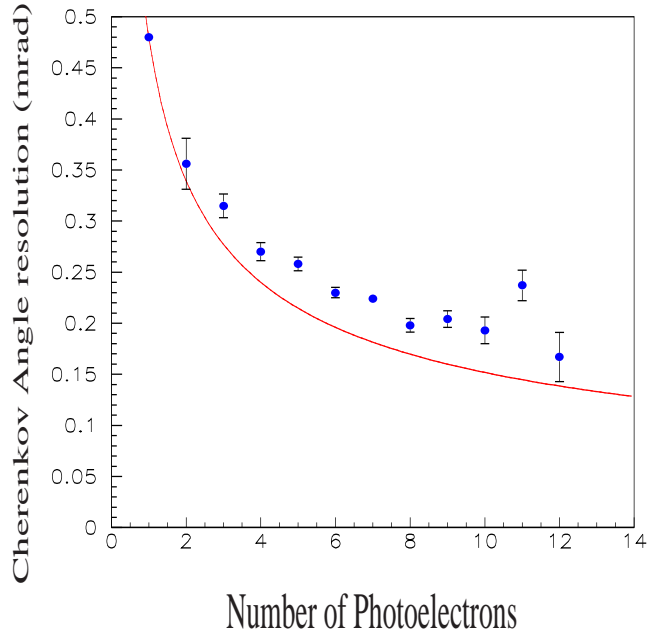,width=0.5\textwidth,height=0.5\textwidth,angle=270.}
\end{center}
\caption{\label{multphot} Cherenkov angular resolution measured using
 61-pixel HPDs vs number of photoelectrons detected in a 
 single particle trigger. The curve is the expectation due to statistical
 errors.}
\end{figure}

     The mean value of the Cherenkov angle from
     all the photoelectron hits in each event is calculated for the
     data from the seven 61-pixel HPDs in 
     \mbox{configuration 1}
     without pyrex filters. The width of
     this distribution versus the number of photoelectrons detected per
     trigger is plotted in Figure \ref {multphot}. For a 
     perfectly aligned system, the width is
     expected to be inversely proportional to the square root of the 
     number of photoelectrons as indicated by the 
     curve. The \mbox{disagreement} between 
     data and simulation is compatible with the residual misalignment
     in the system which is of the order of 0.1 mrad.
   \subsection{Particle Identification}
\begin{figure}
\begin{center}
\epsfig{file=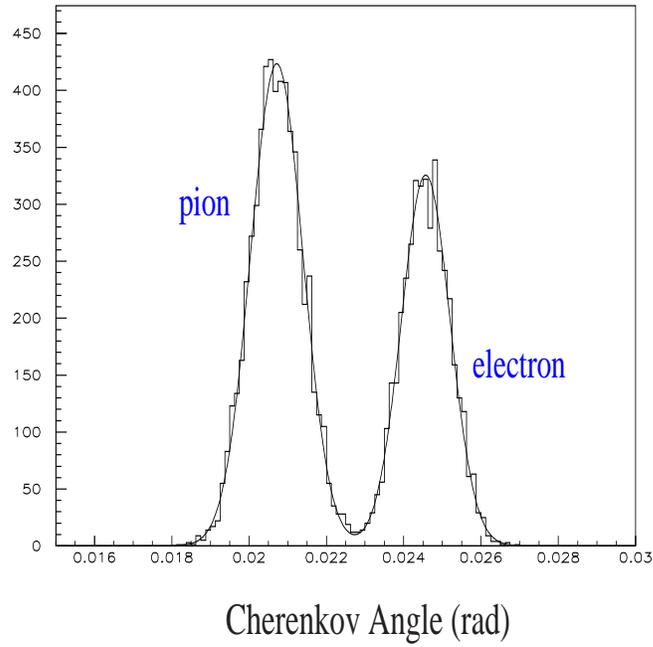,width=0.5\textwidth,height=0.5\textwidth,angle=270.}
\end{center}
\caption{\label{partid} Single photon Cherenkov angle distribution
  for the 2048-pixel HPD without pyrex filter in
  configuration 3 and
  using a 10.4 GeV/{\it c} beam composed of pions and electrons.}
\end{figure}

\begin{figure}
\begin{center}
\epsfig{file=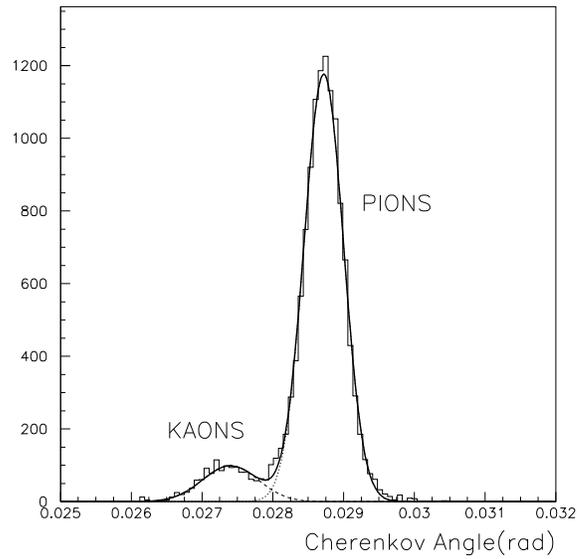,width=0.5\textwidth}
\end{center}
\caption{\label{kpisep} The mean Cherenkov angle from 
  the 61-pixel HPDs without pyrex filter in 
  configuration 1 where the beam was
  a mixture of kaons and pions at 50 GeV/{\it c}.}
\end{figure}

      Figure \ref {partid} shows the Cherenkov angle distribution for
     the 2048-pixel HPD without pyrex filter in 
     configuration 3 where the beam used was a mixture of
     pions and electrons at 10.4 GeV/{\it c}. Good separation is 
     obtained between
     the two particle types.
      Figure \ref {kpisep} shows the plot of the 
      the mean Cherenkov angle calculated from the hits 
      in the 61-pixel HPDs without pyrex filter in 
      configuration 1, 
       where the beam was a mixture of kaons and pions, approximately
       in the ratio 1:9, at 50 GeV/{\it c}.
      Peaks corresponding to the two charged particle types can be
      seen in this figure.
\section {\large Summary and Outlook for the Future}

 The goals set for the RICH2 prototype tests have largely 
 been accomplished. 
The performance of the $CF_{4}$ radiator and the optical 
layout of the RICH2 detector have been tested. 
Photoelectron yields from the prototype HPDs and MAPMTs have been 
measured and found to agree with simulations. A Cherenkov angle precision of
0.35 mrad as assumed in the LHCb technical proposal \cite {lhcb_proposal} 
has been demonstrated with all three photodetectors.

Improvements in the integrated quantum efficiency of both HPDs and
MAPMTs are expected in future devices.
The LHCb RICH detector will require photodetectors with higher active to
total area than those tested here. HPDs with 80\% active area 
\cite {Beaune_conf99} and a lens system for MAPMTs are currently being
developed. These will be tested with LHC compatible readout
(25 ns shaping time) during 1999-2000.
\section {\large Acknowledgements}
 This work has benefited greatly from the technical support 
provided by our colleagues at the institutes
participating in this project. In particular the mirror reflectivity 
and the pyrex transmission were measured by A. Braem. The radiator vessel
extensions were manufactured by D. Clark and I. Clark. The printed
circuits for the MAPMT were designed and assembled by S. Greenwood.
The silicon telescope was provided by 
E. Chesi and J. Seguniot. We also received
valuable advice and assistance from our colleagues in the LHCb
collaboration, in particular from R. Forty, O. Ullaland and T. Ypsilantis.
  
 Finally, we gratefully acknowledge the CERN PS division for providing the 
test beam facilities and the UK Particle Physics and Astronomy Research
Council for the financial support.
%
%
%\clearpage
%


\begin{thebibliography}{99}
\bibitem{lhcb_proposal} LHCb Technical Proposal, CERN/LHCC 98-4.
%
%\bibitem{DEP} Delft Electronische Producten (DEP), The Netherlands.
%
%\bibitem{Hamamatsu} Hamamatsu Photonics , Japan.
%
\bibitem{t_paper2}E. Albrecht $et$ $al.$,  Submitted as an
 \mbox{accompanying} article to Nucl. Instr. and 
   Meth. A in November, 1999.
%
\bibitem{t_paper1}E. Albrecht $et$ $al.$, 
  Nucl. Instr. and Meth. A411(1998) 249.
%
\bibitem{cf4_mea} R. Abjean $et$ $al.$, Nucl. Instr. and Meth A292(1990) 593.
%
\bibitem{viking} O. Toker $et$ $al$., Nucl. Instr. and Meth. A340(1994) 572.
%
\bibitem{lhc1_chip} E. Heijne $et$ $al.$, Nucl. Instr. and Meth 383(1996) 55.
%
%
\bibitem{thiery99} M. Alemi $et$ $al$., Submitted to Nucl. Instr. and Meth.
                     in July, 1999. Also CERN EP/99-110. 
%
\bibitem{bsca_ref} E. Darlington, J. Phys D8 (1975) 85.
%
%
\bibitem{comoproc} M. Alemi $et$ $al$., Nucl. Phys. B 
                   (Proc. Suppl.) 78 (1999) 360-365. 
%
%%\bibitem {bethe} 'Studies in penetration of charged particles in
%     matter', National Academy of Sciences- National Research Council
%     publication 1133, second printing (1967).
\bibitem{Beaune_conf99} E. Albrecht $et$ $al$., Proceedings of the
    second conference on New Developments in Photodetection,Beaune99, 
    France, June 21-25 1999. Submitted to Nucl. Instr. and Meth.

\end{thebibliography}
\end{document}